\documentstyle[prl, multicol, aps, epsf]{revtex}

\begin{document}
\draft

\title
{\large \bf Energy-Dependent LDOS Modulation in Cuprate
Superconductors}
\author{Degang Zhang and C. S. Ting}
\address{Texas Center for Superconductivity and Department of Physics,
University of Houston, Houston, TX 77204, USA}

\maketitle

\begin{abstract}

Motivated by the recent scanning tunneling microscopy (STM)
experiment [J. E. Hoffman, {\it et
al.}, Science {\bf 297}, 1148 (2002)], we investigate the
energy-dependent modulation induced by a weak and extended
defect in a d-wave superconductor. The Fourier component of the
local density of states (LDOS) is calculated up to the first order
in the defect parameters. Our numerically obtained image
maps together with the energy-dependent charge modulation
wave vectors at different dopings exhibit the essential
features as those measured by the experiment.
We also predict new modulation wave vectors in the
first Brillouin zone. Hopefully, they could be verified
by future STM experiments.

\end{abstract}

\begin{multicols}{2}

In the past year, a series of scanning tunneling spectroscopy
(STM) experiments have confirmed the coexistence
of charge modulation and superconductivity in
$Bi_2Sr_2CaCu_2O_{8+\delta}$ [1, 2, 3]. In a magnetic
field, Hoffman {\it et al.} [1] discovered a four cell checkerboard
local density of states (LDOS) modulation localized in a small region
around the vortex core. The field-induced charge modulation
oriented parallel to the $Cu-O$ bond directions
is relatively strong. Subsequently, Howald {\it et al.} [2] also
observed similar checkerboard charge modulation in absence
of magnetic field for a wide range of bias voltages,
but with relatively weak intensity.

Recently Hoffman {\it et al.} [3] investigated the
zero-field charge modulation
by employing high resolution Fourier transform scanning tunneling
spectroscopy. They found that the period of the charge modulation
depends on the energy and doping for the bias voltages  below
the maximum superconducting gap. With the bias voltage
(doping fixed) or doping (bias voltage fixed)
increasing, the LDOS modulation wave vectors oriented parallel to
the $(\pm\pi,0)$ and $(0,\pm\pi)$ directions become shorter
while those oriented parallel to the $(\pm\pi,\pm\pi)$
directions become longer.

A number of theoretical studies on the STM spectra have already
been carried out by several authors [4-7] in attempt to explain
the zero-field LDOS modulation observed by Howald {\it et al.}
[2]. There are also works [8-10] trying to understand the
energy-dependent LDOS modulation observed in the zero-field STM
experiment [3]. In Ref. [8], Wang and Lee proposed that the
experimental observation of Hoffman {\it et al.} [3] is a result
of the quasiparticle interference induced by an impurity with an
on-site potential of moderate strength. However, these
calculations [8-10] seem only able to address a limited portion of
the STM experimental measurements [3], and do not give the
relations among the peaks associated with the modulation wave
vectors, dopings and the bias voltages as those presented in Ref.
[3].

In this paper,  we examine the effects of the scattering of
a quasiparticle by a weak and extended defect or impurity with
both hopping and pairing disorders on the Fourier transform
of the LDOS. Using the first order T-matrix approximation,
we show that our results are consistent with all the
essential features observed in the experiment of
Ref. [3]. In addition, we predict
new modulation wave vectors existed in the
first Brillouin zone, hopefully they could be verified by
future STM experiments.

The Hamiltonian describing the scattering of quasiparticles
from a single defect with local modifications of both
hopping and pairing parameters in a d-wave superconductor
can be written as
$$H=H_{\rm BCS}+H_{\rm imp},$$
$$H_{\rm BCS}=\sum_{{\bf k} \sigma}(\epsilon_{\bf k}-\mu)c^\dag_{{\bf
       k}\sigma}c_{{\bf k}\sigma}+\sum_{\bf k}\Delta_{\bf k}
     (c^\dag_{{\bf k}\uparrow}c^\dag_{-{\bf k}\downarrow}
     +c_{-{\bf k}\downarrow}c_{{\bf k}\uparrow}),$$
$$H_{\rm imp}=\sum_{<i,j>, \sigma}\delta t_{ij}c^\dag_{i\sigma}c_{j\sigma}
            +\sum_{<i,j>}\delta \Delta_{ij}
            (c^\dag_{i\uparrow}c^\dag_{{j}\downarrow}
             +c_{j\downarrow}c_{i\uparrow}) $$
$$~~~~~~~~ +(V_s+V_m)c^\dag_{0\uparrow}c_{0\uparrow}
        +(V_s-V_m)c^\dag_{0\downarrow}c_{0\downarrow}.
\eqno{(1)}$$
Here $\mu$ is the chemical potential to be determined by doping,
$\epsilon_k=t_1({\rm cos}k_x+{\rm
cos}k_y)/2+t_2{\rm cos}k_x{\rm cos}k_y+t_3({\rm cos}2k_x+{\rm
cos}2k_y)/2 +t_4({\rm cos}2k_x{\rm cos}k_y+{\rm cos}k_x{\rm
cos}2k_y)/2+t_5{\rm cos}2k_x{\rm cos}2k_y$, where
$t_{1-5}=-0.5951, 0.1636, -0.0519, -0.1117, 0.0510 (eV)$.
The band parameters are taken from those of Norman {\it et al.}
[11] for $Bi_2Sr_2CaCu_2O_{8+\delta}$, and the lattice
constant $a$ is set as $a=1$.
The order parameter away from the impurity is given by
$\Delta_{\bf k}=\Delta_0({\rm cos}k_x-{\rm cos}k_y)/2$.

Without loss of generality, at the impurity or defect site, we assume
an on-site potential consisting of a nonmagnetic part, $V_s$,
and a magnetic part, $V_m$. The defect also induces a weak local
modification in the hopping, $\delta t$, to the nearest
neighor sites, and a suppression of the superconductivity order parameter
on the four bonds connected to the impurity site,
$\delta \Delta_1$, and on the other twelve bonds
conncted to the nearest neighbor sites, $\delta\Delta_2$.

The Hamiltonian (1) has in fact been successfully applied
by Tang and Flatte [12] to explain
the resonant STM spectra for Ni impurities
in $Bi_2Sr_2CaCu_2O_{8+\delta}$.
In the present situation, no resonances in
LDOS have been observed in the recent STM experiments [2, 3]. So it is
reasonable to assume that the on-site potentials
($V_s$ and $V_m$) and the modifications in hopping and
pairing parameters ($\delta t$, $\delta \Delta_1$ and
$\delta \Delta_2$) are all weak and have approximately
the same order of magnitude. In order to compare with
the measurements in the STM experiment [3],
we investigate three different hole doping cases:
underdoping (10 percent) with $\Delta_0=50$ meV,
optimal doping (15 percent) with $\Delta_0=44$ meV,
and overdoping (17 percent) with $\Delta_0=37$ meV. Here we
shall not discuss the issue as to why the underdoped case
has a higher $\Delta_0$ value, and simply accept it as an
experimental fact [3].

Our method to solve the Hamiltonian (1) and to obtain the LDOS is the standard
Bogoliubov transformation plus Green's
function technique. When $\delta t, \delta\Delta_1,
\delta\Delta_2, V_s$ and $V_m$
are all small, keeping the leading term in the T-matrix approach
should be good approximation. The Fourier component of the LDOS
obtained in such an approximation can be shown to have
the following form
$$ \rho_{\bf q}(\omega)=-\frac{2}{\pi N^2}\sum_{\bf k}\sum_{
   \nu,\nu^\prime=0,1}\{[2\delta t A({\bf k}, {\bf q})+V_s]
   \alpha_{\nu\nu^\prime}({\bf k}, {\bf q})$$
$$ +2[\delta\Delta_1 B({\bf k}, {\bf q})+\delta\Delta_2 C({\bf k}, {\bf q})]
    \beta_{\nu\nu^\prime}({\bf k}, {\bf q})\}$$
$$   \times {\rm Im}[
    G^0_{{\bf k}\nu}(i\omega_n)G^0_{{\bf k}+{\bf q}\nu^\prime}
    (i\omega_n)]|_{i\omega_n\rightarrow \omega+i0^+},
\eqno{(2)}$$
where $N$ is the number of sites in the lattice, $A({\bf k}, {\bf q})
=cosk_x+cosk_y+cos(k_x+q_x)+cos(k_y+q_y)$, $B({\bf k}, {\bf q})
=cosk_x-cosk_y+cos(k_x+q_x)-cos(k_y+q_y)$, $C({\bf k}, {\bf q})
=cos(k_x+2q_x)-cos(k_y+q_x+q_y)-cos(k_y-q_x+q_y)+cos(k_x+q_x+q_y)
+cos(k_x+q_x-q_y)-cos(k_y+2q_y)+cos(k_x-q_x)+cos(k_x-q_y)-
cos(k_y-q_x)-cos(k_y-q_y)+cos(k_x+q_y)-cos(k_y+q_x)$,
$\alpha_{\nu\nu^\prime}({\bf k}, {\bf q})=\xi^2_{{\bf k}\nu}
\xi^2_{{\bf k}+{\bf q}\nu^\prime}-(-1)^{\nu+\nu^\prime}
\xi_{{\bf k}\nu}\xi_{{\bf k}\nu+1}\xi_{{\bf k}+{\bf q}\nu^\prime}
\xi_{{\bf k}+{\bf q}\nu^\prime+1}$, $\beta_{\nu\nu^\prime}({\bf k},
{\bf q})=(-1)^\nu\xi_{{\bf k}\nu}\xi_{{\bf k}\nu+1}
\xi^2_{{\bf k}+{\bf q}\nu^\prime}+(-1)^{\nu^\prime}\xi^2_{{\bf k}\nu}
\xi_{{\bf k}+{\bf q}\nu^\prime}\xi_{{\bf k}+{\bf q}\nu^\prime+1}$,
$G^0_{{\bf k}\nu}(i\omega_n)=1/[i\omega_n-(-1)^\nu E_{{\bf k}}]$
is the bare Green's function, $E_{{\bf k}}=\sqrt{(\epsilon_{\bf
k}-\mu)^2 +\Delta^2_{{\bf k}}}$, $\xi^2_{{\bf k}\nu}=[1+(-1)^\nu
(\epsilon_{\bf k}-\mu)/E_{{\bf k}}]/2$, and $\xi_{{\bf k}0}
\xi_{{\bf k}1}=\Delta_{{\bf k}}/(2E_{{\bf k}})$.

It is noted that $V_m$ is absent from Eq. (2) because
there is no first order contribution from
the magnetic potential.
In the present study, we base our numerical calculation on
a finite lattice of $800\times 800$ lattice sites with
the defect at the center. For simplicity, we choose
$2\delta t=V_s=-2\delta\Delta_1=-4\delta\Delta_2$,
and assume that all these parameters are small such that
the first order T-matrix approximation is valid. The
advantage of this first order approximation is that
the states of quasiparticle before and after scattering
from the defect are clearly distinguishable.
In our calculation, we also
introduce a finite lifetime
broadening $\gamma=2$ meV to the quasiparticle Green's
function to smooth our data points by replacing $\omega +i0^+$
with $\omega +i\gamma$ in Eq.(2).

It is well-known that the quasiparticles in a d-wave superconductor
are Bloch wave states. In the presence of electron-impurity interactions,
elastic scattering mixes those eigenstates of the quasiparticle
with the same energy but different momentum. For example,
if a quasiparticle with energy $E_{\bf k}$ is excited at the
point O in Fig. 1(a), after being scattered by the defect,
the quasiparticle energy become $E_{\bf k+q}$ ($\equiv E_{\bf k}$).
There are 6 nonequivalent ${\bf q}$ vectors
as shown in Fig. 1(a). The variation in the
magnitudes of these ${\bf q}$ vectors with $\omega$ would
lead to energy-dependent LDOS modulations.

According to Eq. (2), we plot the image map of the Fourier component
of LDOS in Fig. 1(b) for the optimally doped case at fixed
$\omega=-16$ meV in the first Brillouin zone. In Fig. 1(b)
we are able to clearly identify four of the six ${\bf q}$
vectors as shown in Fig. 1(a) from the positions of
the peaks (or local maxima). The four peaks with relatively weak
intensity along $(\pm \pi, 0)$,
$(0, \pm \pi)$ directions and the four peaks with relatively strong intensity
along $(\pm \pi, \pm \pi)$
directions as shown in Fig. 1(b) are respectively related to
${\bf q}_A$ and ${\bf q}_B$ in Fig. 1(a), and both of them
have been observed in
Ref. [3]. In addition, we predict that there are another
eight weak peaks corresponding to ${\bf q}_C$. At the four corners
of the first Brillouin zone,
there are four extended weak peak arcs
generated from the quasiparticle interference between the banana-
shaped equal-energy contours of the diagonal Fermi surfaces
connected by ${\bf q}_D$.
Since the peaks at ${\bf q}_B$
have the highest intensity, we expect that the real space LDOS image
at $\omega=-16$ meV should have a checkerboard pattern oriented
along $45^0$ to the $Cu-O$ bonds with a period close to $5a$.

In Fig. 2(a) to 2(c), the q-space LDOS maps
for the optimally doped case are also presented
at $\omega=0, -12, -20$ (meV) in the first Brillouin zone.
We noticed that the detailed image of the map depends on the
magnitude of $\omega$, when it increases,
$|{\bf q}_A|$ becomes shorter while $|{\bf q}_B|$,
$|{\bf q}_C|$ and $|{\bf q}_D|$ become longer.
In Fig. 2(c) with $\omega=-20$ meV, the intensity of the peaks
at ${\bf q}_A$ is catching up with that at ${\bf q}_B$
as compared with the case of $\omega=-16$ meV.
And it is expected to
become dominant at a larger $|{\omega}|$, there, the real
space LDOS image would have a checkerboard pattern oriented
along the $Cu-O$ bonds with a period close to $4a$.
This is consistent with the experiment in Ref. [3].

It needs to be mentioned that the peaks
corresponding to the remaining two
modulation wave vectors ${\bf q}_E$ and ${\bf q}_F$
[see Fig. 1(a)] are clearly missing in these maps
[see Fig. 1(b), and Fig. 2(a) to 2(c)].
At $\omega=0$ meV, ${\bf q}_D$ and ${\bf q}_E$
are equivalent and a small arc peak is created at
them, but the resolution of the map is not clear
enough to show the existence of ${\bf q}_F$
(i.e. ${\bf q}_A$ and ${\bf q}_C$).
For $\omega=-12, -16, -20$ (meV),
both of these wave vectors do not appear in the
first Brillouin zone, but they may yield local peaks in the
second Brillouin zone. So far the experiments have
not yet been performed in this region, their effects will not
be considered in the present study.
But we do expect that the peaks associated with these
wave vectors would show up when the boundary
of our calculation is expanded beyond that of the first
Brillouin zone.
Here we would like to point out that although our image
maps at $\omega=-12$ meV is similar to corresponding one
in Ref. [8], there still exist some fundamental differences
between these two results. For example, the bright spots
(the dominant peaks)
at points $(\pm\pi,\pm\pi)$ appeared in the image maps of Ref. [8]
close to $\omega=-20$ meV did not show up in our maps.
If this prediction in Ref. [8] is true, the real space LDOS
would have a checkerboard pattern with the period $1.414a$
oriented along $45^0$ to the $Cu-O$ bonds, which seems to be
contradicting to the experimental observation [3].

At $\omega=12, 16, 20$ (meV),
the maps of the Fourier component of LDOS
for the optimally doped case are shown in
Fig. 2(d) to (f). The modulation wave vectors are
identical to those for the negative $\omega$. But the
Fourier component of LDOS at ${\bf q}_B$, ${\bf q}_C$ and
${\bf q}_D$ shows a local minimum intensity. This is primarily
due to the destructive interference between the opposite
phases carried by the quasiparticles before and after
scattering. This destructive interference has also been
observed at $\omega=16, 22$ (meV) in the STM experiment of Ref.
[3] (see Fig. 2C and 2D there). Obviously,
the wave vectors at extreme
(maximum or minimum) values of $\rho_{\bf q}(\omega)$
for the positive $\omega$ coincide with those in
Fig. 1(b), Fig. 2(b) and 2(c). In addition,
we also found that in higher energy
region close to the maximum of the superconductivity gap,
our image maps become blurred. The reason is not clear to us,
but it might be due to many competing modulation wave vectors
gaining strengths on the same banana shaped equal energy
contour of the Fermi surface.

In order to compare with the experimental curves in
Ref. [3], Fig. 3 shows the Fourier component of LDOS respectively
along $(\pi, 0)$ [Fig. 3(a)] and $(\pi, \pi)$ [Fig. 3(b)] directions
as a function of $|{\bf q}|$  up to  $|{\bf q}|=0.4$
(in unit of $2\pi$) for the optimally doped sample.
Along $(\pi, 0)$ direction, the peak of $\rho_{\bf q}(\omega)$
associated with ${\bf q}_A$
moves slowly towards the origin when energy increases.
In contrast, along $(\pi, \pi)$ direction, the peak of
$\rho_{\bf q}(\omega)$ associated with ${\bf q}_B$
moves rapidly  away from the origin.
This energy-dependent position of the peaks
at ${\bf q}_A$ or ${\bf q}_B$
is responsible for the checkerboard pattern
with an energy-dependent period as observed in Ref. [3].
In the presence of other competing modulation wave
vectors at higher energy, the LDOS image pattern could be
dramatically modified.

We also study the effects of doping on $\rho_{\bf q}(\omega)$.
Fig. 4 shows the peak positions
at $|{\bf q}_{A}|$ along $(\pi, 0)$ direction [Fig. 4(a)] and
at $|{\bf q}_{B}|$ along $(\pi, \pi)$ direction [Fig. 4(b)]
as a function of ${\omega}$
at three different dopings. As energy is fixed,
$|{\bf q}_{A}|$ becomes smaller while
$|{\bf q}_{B}|$ becomes larger as the doping is increased.
It is apparent that the results in Fig. 3 and 4 are in qualitative
agreement with those experimental curves in Ref. [3].

In summary, we have studied the energy-dependent LDOS modulations
in d-wave superconductors with the presence of an extended
defect using the Bogoliubov transformation plus the Green's
function approach.
The changes in pairing order parameter and hopping terms
due to such an extended defect generate some ${\bf k}$
and ${\bf q}$ dependent terms as shown in Eq. (2),
which seem to be  essential for obtaining the curves
in Figs. 3 and 4. In addition, we also discover
new modulation wave vectors ${\bf q}_C$ and ${\bf q}_D$
in the first Brillouin zone. Hopefully, their effects
could be observed in future experiments.
Since the effects of the modulation wave vectors in
the second Brillouin zone have not been carefully
examined, we are not able to obtain the complete LDOS
image in real space. This is an important problem
and should constitute a subject for future investigation.
We also note that using the triangular relation ${\bf q}_A
+{\bf q}_B={\bf q}_C$, the Fermi surface and the energy
gap could be mapped out.

We wish to thank Prof. J. C. Davis for sending us Ref. [3]
before its publication, and H. Y. Chen, M. Shaw, Y. Chen
and Prof. S. H. Pan for useful discussions.
This work has been supported by the Texas Center for Superconductivity
at the University of Houston and by the Robert A. Welch Foundation.

\begin{figure}
\epsfxsize=1cm \vspace{2cm} \narrowtext \caption {(a) Schematic
Fermi surface of high-$T_C$ cuprate superconductor. The modulation
wave vectors connecting different points of the Fermi surface
with the same energy gap are shown. (b) The Fourier component
of LDOS $\rho_{\bf q}(\omega)$
for the optimally doped case at $\omega=-16$ meV
in the first Brillouin zone (i.e. $-\pi<q_x, q_y\leq \pi$). }
\end{figure}

\begin{figure}
\caption {
The Fourier component of LDOS $\rho_{\bf q}(\omega)$
for the optimally doped case
at different energy shown in each panel in the first Brillouin zone.}
\end{figure}

\begin{figure}
\caption {$\rho_{\bf q}(\omega)$
versus $|{\bf q}|$ for the optimally doped case along
(a) the $(\pi, 0)$ direction and (b) the
$(\pi, \pi)$ direction are shown for seven quasiparticle energies.
The data are shifted vertically relative to each other by
$0.5$ unit for clarity.}
\end{figure}

\begin{figure}
\caption {The modulation
wave vectors versus energy along (a) the $(\pi, 0)$ direction and
(b) the $(\pi, \pi)$ direction are plotted for
under-, optimally and over-doped cases.}
\end{figure}

\end{multicols}

\end{document}